# A green solvent system for precursor phase-engineered sequential deposition of stable formamidinium lead triiodide for perovskite solar cells

Benjamin M. Gallant[1,2], Philippe Holzhey[1], Joel A. Smith[1], Saqlain Choudhary[1], Karim A. Elmestekawy[1], Pietro Caprioglio[1], Igal Levine[3], Alex Sheader[1], Fengning Yang[1], Daniel T.W. Toolan[4,5], Rachel C. Kilbride[5], Augustin K.A. Zaininger[1], James M. Ball[1], M. Greyson Christoforo[1], Nakita Noel[1], Laura M. Herz[1,6], Dominik J. Kubicki[2], Henry J. Snaith[1]*

[1] Clarendon Laboratory, Department of Physics, University of Oxford, Parks Road, Oxford, OX1 3PU, United Kingdom
[2] School of Chemistry, University of Birmingham, B15 2TT, Birmingham, UK
[3] Division Solar Energy, Helmholtz-Zentrum Berlin für Materialien und Energie GmbH, Berlin, 12489, Germany Solar Energy Division
[4] Department of Materials, University of Manchester, Manchester, M13 9PL, UK
[5] Department of Chemistry, University of Sheffield, Sheffield, S3 7HF, UK
[6] Institute for Advanced Study, TU Munich, Lichtenbergstr. 2a, 85748 Garching, Germany

## Abstract

Perovskite solar cells (PSCs) offer an efficient, inexpensive alternative to current photovoltaic technologies, with the potential for manufacture via high-throughput coating methods. However, challenges for commercial-scale solution-processing of metal-halide perovskites include the use of harmful solvents, the expense of maintaining controlled atmospheric conditions, and the inherent instabilities of PSCs under operation. Here, we address these challenges by introducing a high volatility, low toxicity, biorenewable solvent system to fabricate a range of 2D perovskites, which highly effective precursor phases for subsequent transformation to α-formamidinium lead triiodide (α-$FAPbI_3$), fully processed under ambient conditions. PSCs utilising our α-$FAPbI_3$ reproducibly show remarkable stability under illumination and elevated temperature (ISOS-L-2) and “damp heat” (ISOS-D-3) stressing, surpassing other state-of-the-art perovskite compositions. We determine that this enhancement is a consequence of the 2D precursor phase crystallisation route, which simultaneously avoids retention of residual low-volatility solvents (such as DMF and DMSO) and reduces the rate of degradation of $FA^+$ in the material. Our findings highlight both the critical role of the initial crystallisation process in determining the operational stability of perovskite materials, and that neat $FA^+$-based perovskites can be competitively stable despite the inherent metastability of the α-phase.

## Introduction

Scalable, reproducible and controlled crystallisation of highly stable metal halide perovskite thin films is crucial to the commercial implementation of this promising class of semiconductor materials in optoelectronic applications. Ideally, solution-processed perovskite deposition should employ inks based on non-toxic and volatile solvents and should result in the controlled formation of a well-mixed and low defect perovskite phase that – crucially – possesses long term operational stability under external stimuli such as heat, light and moisture[1–3].

Significant attention has been paid to the relationship between halide perovskite composition ($ABX_3$ – where A is an organic or alkali-metal cation, B is a lead or tin cation, and X are halide anions) and operational stability. Exchanging the archetypical A-site methylammonium ($MA^+$) cation for formamidinium ($FA^+$) significantly increases thermal and photochemical stability, but introduces a phase instability. The photoactive α-$FAPbI_3$ phase is only metastable at room temperature, and the transition to the yellow non-perovskite δ-phase is accelerated by the presence of moisture[4]. This instability has been circumvented by alloying $FA^+$ with caesium cations ($Cs^+$)[2]. However, A-site inhomogeneity has been linked to both poor optoelectronic performance[5] and other long-term degradation pathways, such as the emergence of $Cs^+$-rich impurity phases[6,7]. Further, of all the elements employed in halide perovskites, the scarcity of Cs poses the greatest challenge in availability for TW scale PV production[8] and thus routes realising stable neat-$FA^+$ perovskites are greatly advantageous.

However, despite the emphasis on $ABX_3$ composition as the key factor controlling perovskite operational stability, properties such as the types[9] and density[10] of defects, impurity phases[11], compositional inhomogeneity[12], residual solvents[13–15] and residual strain[16] have also been shown to influence long-term stability. All such properties are imparted into the perovskite thin layer during crystallisation of the $ABX_3$ structure, which can proceed via a range of intermediates or precursor phases, including solvate[17–19] and polytype[20,21] phases and phases incorporating sacrificial ions[22]. Given the origin of these properties, it is imperative that the relationship between crystallisation pathway, precursor phase formation and resulting $ABX_3$ perovskite stability be investigated and understood more widely.

Despite substantial research over the past decade[23,24], highly toxic *N*,*N*-dimethylformamide (DMF) remains ubiquitous in halide perovskite inks, while operational instability of perovskites-based optoelectronics remains a key barrier to their wider application[1]. Here we simultaneously combat the challenges of perovskite ink toxicity, processing-induced instability and compositional heterogeneity. By developing a precursor ink that utilises a highly volatile, low toxicity solvent mixture we replace conventional precursor phases with a solid-state 2D perovskite material. By exchanging the organic cations in the 2D precursor phase with $FA^+$, we enable the subsequent crystallisation of α-$FAPbI_3$ perovskite. We not only find that our α-$FAPbI_3$ performs effectively in PSCs, but that 2D-precursor phase growth produces α-$FAPbI_3$ with substantially improved ambient, thermal and photostability in comparison to any other perovskite composition or processing route that we fabricate herein[2,3]. We are able to fabricate neat $FAPbI_3$ thin films that are stable for >3,000 hours under harsh heat, light and moisture conditions, and achieve a promising champion lifetime to 80% of initial performance ($t_{80}$) of 800 hours and no degradation ($t_{100}$) for more than 1,930 hours under ISOS-L-2 (85 °C, 1-sun equivalent) and ISOS-D-3 (85 °C, 85% relative humidity), respectively, when integrated into PSCs. Our work enables us to distinguish how different processing variables impact perovskite stability and demonstrate why conventional solution processing routes are not only problematic for toxicity, but are fundamentally linked to halide perovskite instability.

## A biorenewable and low toxicity solvent system

As solution processing at industrial scale dictates that the majority of solvents employed must be volatilised into the atmosphere[25], the hazard of employing toxic solvents in PSC production at increasingly large scale are substantial. In light of this, several DMF-free solvent systems have been established for perovskite solution processing, notably based on *N*,*N*-dimethylacetamide (DMAc)[26], γ-butyrolactone (GBL)[27], 2-methoxyethanol (2-ME)[28] and acetonitrile (ACN) in combination with methylamine gas[29]. However, as **Fig. 1a** shows, DMAc, 2-ME and ACN do not provide the desired reduction in toxicity, while GBL is an illegal narcotic on account of its metabolite, γ-hydroxybutyrate (GHB)[30]. Furthermore, the use of MA gas as a co-solvent, poses further problems of volatility, where escape of the MA gas can render the solution unstable and pose a toxicology risk itself. Moreover, while dimethyl sulfoxide (DMSO) – the typical companion to DMF in perovskite solution processing – is non-toxic as a neat liquid, in the context of lead salt solvation the proficiency of DMSO as a percutaneous absorption (skin-penetration) enhancer[31,32] is of acute concern for at-scale processing (**Fig. 1b**).

Here we investigate and develop a solvent system consisting of biorenewable cyclic ethers and alkylamine liquids, in which the latter serves both as a coordinating co-solvent and subsequently as a component within the 2D perovskite precursor phase. This system can be considered part of a family of solvent mixtures all comprising poorly $Pb^{2+}$-coordinating solvents as the majority component alongside strongly $Pb^{2+}$-coordinating solvents[28]. Here cyclic ethers take on the role of the poorly coordinating solvent while butylamine (BA), which is liquid at room temperature, acts as the strongly coordinating solvent. We find that both tetrahydrofuran (THF) and 2-methyltetrahydrofuran (MeTHF) are suitable poorly-coordinating solvents, however we focus on the latter on account of its biorenewable production, significantly lower toxicity and favourable vapour pressure for rapid at-scale solution processing. Vidal, et al.[25], recently highlighted that THF's high vapour pressure renders its lifetime environmental impact relatively large compared with less volatile perovskite solvents due to additional greenhouse gas emissions resulting from energy-intensive solvent recapture. However, a lifecycle analysis of MeTHF conducted by Slater, et al.[33], demonstrated a 97 % reduction in lifetime emissions associated with the use of MeTHF as compared to THF, due mostly to its biorenewable production and significantly lower vapour pressure. Moreover, MeTHF has recently been recommended by both the European Medicines Agency (EMA) and the Food and Drug Administration (FDA) as a Class 3 solvent ("solvents with low toxic potential") with a permitted daily exposure limit of 50 mg $day^{-1}$ [30]. This exposure limit is much higher than that of THF (7.2 mg $day^{-1}$), which Vidal, et al.[25], demonstrate already possesses significant toxicological advantages over other common perovskite solvents. Furthermore, as solution processing demands that the majority of solvents employed are volatilised, the other safety-critical parameters for at-scale processing are the no observed adverse effect concentration (NOAEC) via inhalation and the lower flammability limit, rather than more conventional measures of toxicity, such as the lethal dose (LD50) or concentration (LC50). Based on the European Chemicals Agency's (ECHA) NOAECs (exposure via inhalation) for MeTHF and BA we calculate a NOAEC for our mixed solvent system (9:1 MeTHF:BA) of 1,150 mg $m^{-3}$, a 185-fold increase in comparison to DMF (6.2 mg $m^{-3}$) (**Fig. 1a**). The lower flammability limit of all solvents commonly used for perovskite processing are comparable. Full discussion of these parameters is offered in **Supplementary Note 2**. Besides these, the critical advantage of both THF and MeTHF over the previously reported ACN systems is that use of highly volatile MA gas is precluded. Use of a liquid – rather than gas phase – amine minimises the complexity of precursor preparation and handling, improving scalability and ensuring reproducible solution composition.

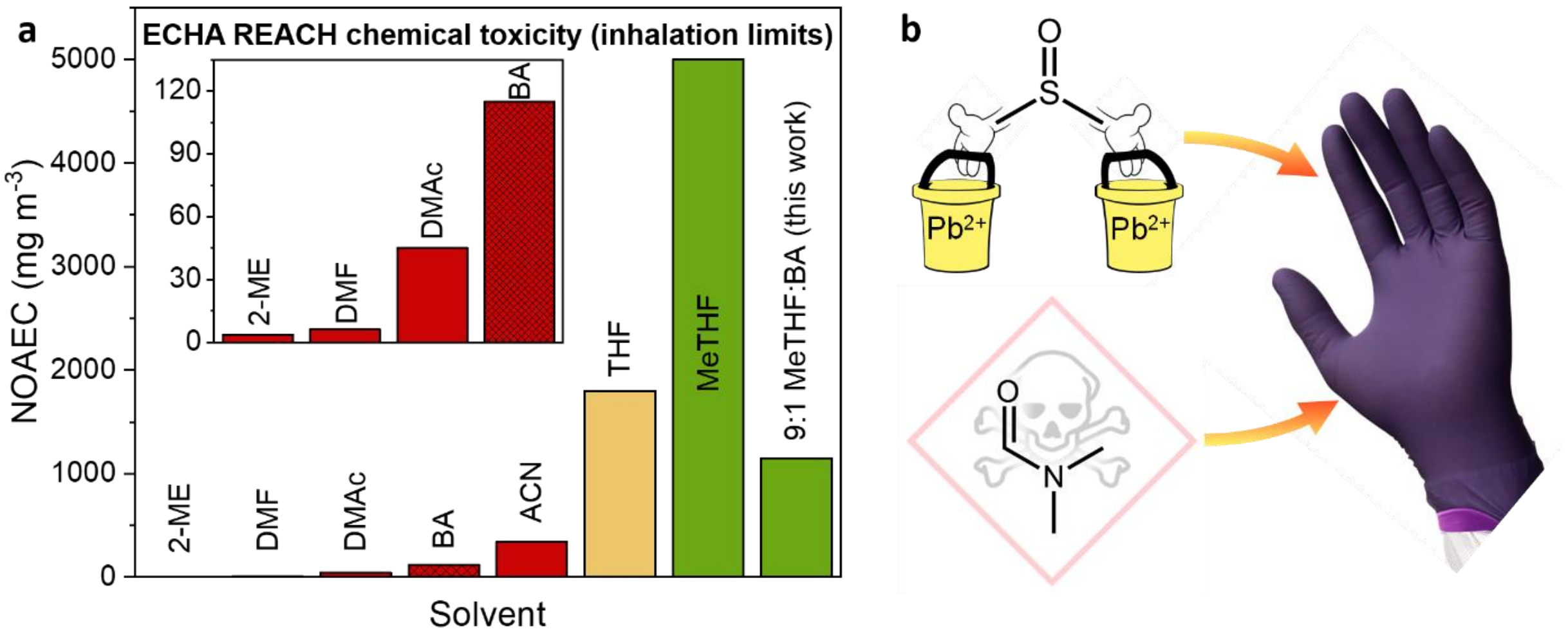

**Fig. 1 | Scalable processing compatibility. a**, Preferential properties of THF and MeTHF as solvents in industrial processes where vapour phase exposure is unavoidable. Solvents shown in green are produced industrially from biorenewable sources, in yellow may be produced biorenewably but typically are not, and in red cannot be produced from renewable sources. No observed adverse effect concentrations (NOAECs) are taken from the European Chemicals Agency (ECHA) database and are calculated for workplace (except BA, where the value for general population is used) exposure via inhalation of solvent vapours. **b**, Schematic demonstrating toxicity dangers concerning lead salt-containing DMF (carcinogen) and DMSO (skin-penetration enhancer) solvents.

**2D precursor phase-growth of 3D perovskites**

Notwithstanding the volatility of MA, amines hold their own critical advantage against other Lewis basic $Pb^{2+}$-coordinating solvents; they also behave as Brønsted-Lowry bases in precursor solutions. In our precursor ink the inclusion of BA alongside methylammonium iodide (MAI) leads to a Brønsted-Lowry $H^+$ exchange in solution and the *in-situ* production of *n*-butylammonium ($BA^+$). We dissolve MAI and lead (II) iodide ($PbI_2$) in our MeTHF:BA solvent system and cast a thin film by spin-coating and drying at 70 °C for ten minutes. Incorporation of both $BA^+$ and methylammonium ($MA^+$) within the perovskite structure leads to highly crystalline 2D Ruddlesden-Popper phase perovskites (RPPs) of the $(BA^+)_2(MA^+)_{n-1}Pb_nI_{3n+1}$ family, where $BA^+$ serves as the large A-site cation (A') and $MA^+$ as the small A-site cation (A)[34]. The *n* described by this empirical formula is often used to differentiate RPP phases by the number (n) of lead-halide octahedra ($[PbI_6]^{4-}$) layers interspersed between layers of the large A' cations. In steady-state photoluminescence (PL), X-ray diffraction (XRD) and absorbance measurements (**Supplementary Fig. 2-4**) we observe multiple peaks consistent with the presence of multiple RPP phases. By varying BA:$MAPbI_3$ precursor stoichiometry ($R_{BA\text{-}MA+}$) we are able to control the average RPP phase composition, denoted <*n*> (**Supplementary Note 3**).

Although 2D RPPs have been utilised as photo-absorbers in efficient PSCs, these typically display non-ideal bandgaps for use in single-junction photovoltaics[35]. Instead, we propose an alternative application of 2D mixed-phase RPPs; as solid-state precursor phases, or intermediates, in a solution-processed sequential deposition of 3D $ABX_3$ perovskites[36,37]. This strategy allows fabrication of α-$FAPbI_3$ by our solvent system but avoids the reported reactivity of amines with $FA^+$ in solution[38,39].

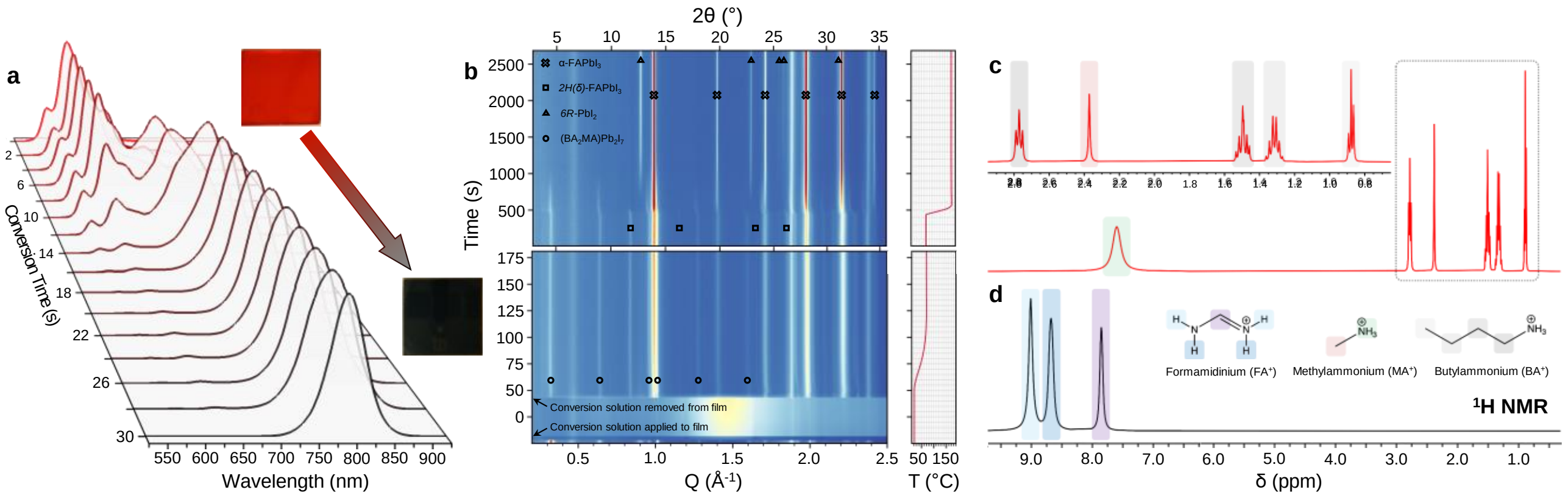

**Fig. 2 | Conversion of 2D precursor phase to 3D α-$FAPbI_3$ perovskite. a**, Tracking photoluminescence (normalised) of optimised 2D precursor phase ($R_{BA\text{-}MA+}$ = 1.5) converting to 3D α-$FAPbI_3$ during soaking in "conversion solution" (FAI solution in *n*-butanol). Unconverted 2D precursor phase shown at t = 0 s. Insets: images of 2D precursor phase (red) and 3D (black) perovskite thin films. **b**, *In-situ* grazing-incidence wide-angle X-ray scattering (GIWAXS) tracking evolution of crystalline phases present as 2D perovskite is converted to α-$FAPbI_3$. $^1$H solution nuclear magnetic resonance (NMR) spectra of (**c**) 2D precursor phase ($R_{BA\text{-}MA+}$ = 1.5) thin films (inset shows expanded region of spectrum corresponding to methyl and methylene signals, highlighted in dashed box) and (**d**) α-$FAPbI_3$ thin films dissolved into DMSO-$d_6$.

To convert the 2D precursor phase into α-$FAPbI_3$, we dispense a solution of formamidinium iodide (FAI) in *n*-butanol ("conversion solution") on top of the substrate and allow conversion *via* cation exchange to occur over a controlled time before spin-coating. In **Fig. 2a** we present the PL spectra of the precursor RPP layer as a function of conversion time. **Supplementary Fig. 8** tracks the corresponding evolution of absorbance during conversion. While $FA^+$ cations intercalate into the 2D material, we observe a gradual reduction in the PL intensity at the original emission wavelengths and the emergence of longer wavelength emission features. We interpret this to be due to $FA^+$ exchanging with $MA^+$ and $BA^+$ and at first increasing the <*n*>-value in the now "triple cation" RPP perovskite, before the resultant film becomes predominantly $FAPbI_3$, or a mixed $FA_xMA_{1-x}PbI_3$ 3D perovskite, as judged by the emission peak at 780 nm. After spin-coating to remove excess conversion solution, subsequent thermal annealing at low (70 °C, 10 minutes) and then high (180 °C, 30 minutes) temperature completes conversion to the 3D perovskite, with the high temperature step selected with the intention of driving out remnant $BA^+$ and $MA^+$. The final emission peak of the annealed perovskite film is centred at 821 nm. **Fig. 2b** shows the same process tracked by *in-situ* GIWAXS, highlighting the evolution of crystalline domains during conversion and subsequent annealing. After solution conversion we find that only a small quantity of the 2D precursor phase is retained, with the majority having been converted directly to 3D perovskite. Reflections consistent with a very small quantity of the preferred RT phase (*2H*-$FAPbI_3$) are observed, however these rapidly disappear along with remnant 2D intermediate upon high temperature curing. The direct conversion from 2D to 3D perovskite indicated by these combined *in-situ* GIWAXS and PL results indicate that the extended corner-sharing [$PbI_6$] octahedral network present in the precursor phase is retained into the 3D perovskite.

During *in-situ* GIWAXS measurement of the first five minutes of high temperature annealing, reflections corresponding to *6R*-$PbI_2$ gradually appear, but the intensity and position of these peaks stabilises – along with those of α-$FAPbI_3$ – consistent with complete volatilisation of small quantities of remnant $MA^+$ from the 3D perovskite phase. As $MA^+$ is present in the precursor phase, it is important to identify if any residual $MA^+$ is left in the final annealed film, since if it were this could contribute to the $ABX_3$ phase stabilisation. Furthermore, previous work has indicated the phase stability of α-$FAPbI_3$ is increased with the incorporation of some 2D phases[40], and hence determining if any residual $BA^+$ is present is also important. Ex-situ XRD characterisation of the fully-annealed 3D perovskite shows a reflection at 2θ = 13.92° (**Supplementary Figure 15a**), consistent with that predicted for the (100) reflection of α-$FAPbI_3$ at 298 K (CCDC: 2243718; 2θ = 13.924°)[41]. We also perform $^1H$ solution NMR spectroscopy measurements of the as-annealed 3D perovskite films dissolved in DMSO-$d_6$ (**Fig. 2d-e**), which confirm the absence of residual $MA^+$. The clear resolution of $^{13}C$ satellite signals ($^1J_{1H\text{-}13C}$ coupling) indicates sensitivity to trace $MA^+$ or $BA^+$ as low as ~0.1 mol% (**Supplementary Fig. 11**), placing an upper limit residual $MA^+$ or $BA^+$ in our material. Further, in **Supplementary Fig. 12** we show scanning electron diffraction (SED) measurements of the final annealed precursor phase-engineered 3D perovskite films. SED measurements allow detection and structural indexing of nanoscale crystalline domains that are too small for detection by bulk diffraction methods[42]. In these images we cannot find any diffraction spots which correspond to the unit cell dimensions expected from 2D perovskite phases. Thus, although this does not prove that there are no 2D domains present, we see no evidence that there are. From here on, we refer to our annealed 3D perovskite as α-$FAPbI_3$.

To further investigate the mechanism of 2D precursor phase-conversion we assess the impact of precursor composition on the resulting 3D perovskite. Scanning electron microscopy (SEM) images of α-$FAPbI_3$ processed *via* 2D precursor films of varying <*n*> reveal a wide variety of microstructures. Use of a high $R_{BA\text{-}MA+}$ (BA:$MA^+$ precursor solution stoichiometry) 2D precursor phase (low <*n*>) results in a mesoporous structure in the 3D perovskite film (**Fig. 3a-b**), and even macroscopic cracking of the thin-films (**Fig. 3a**, inset). XRD analysis of α-$FAPbI_3$ processed from intermediate $R_{BA\text{-}MA+}$ ratios reveals how gradual reduction of $R_{BA\text{-}MA+}$ (toward higher <*n*>) leads to a pronounced increase in lead (II) iodide (*6R*-$PbI_2$) content in the as annealed films (**Fig. 3i**, **Supplementary Fig. 14**). We observe a consistent trend with the appearance of hexagonal *6R*-$PbI_2$ platelets, which appear brighter in the SEM images (**Fig. 3c-g**)[43]. Conversion from high-*n* 2D precursor phase films also leads to mesoscopic porosity visible *via* SEM (**Fig. 3h**). However, in contrast to low-*n* precursor phases, XRD analysis of high-*n*-converted 3D materials demonstrates significant *6R*-$PbI_2$ content (**Fig. 3i**). We rationalise these results by considering the careful balance of factors that control optimum <*n*> for 2D precursor phase conversion and identify a 'Goldilocks region' in which optimal composition and morphology are achieved. We discuss our full mechanistic rationalisation of the conversion process in detail in **Supplementary Note 4**. Here we summarise our findings.

For effective conversion of the precursor phase, we hypothesised that two processes should be optimised. (1) Facile intercalation of $FA^+$ cations into, and expulsion of $MA^+$ and $BA^+$ from the precursor phase must occur. (2) Volume contraction and structural rearrangement required upon conversion from the 2D perovskite to α-$FAPbI_3$ must be minimised. Both these processes are, in turn, controlled by two properties of the 2D RPP precursor phase; spacer molecule channel density, and the thickness of the corner-sharing haloplumbate, $[PbI_6]_n$, layers in the 2D precursor phase. Both these properties are determined by the <*n*> value.

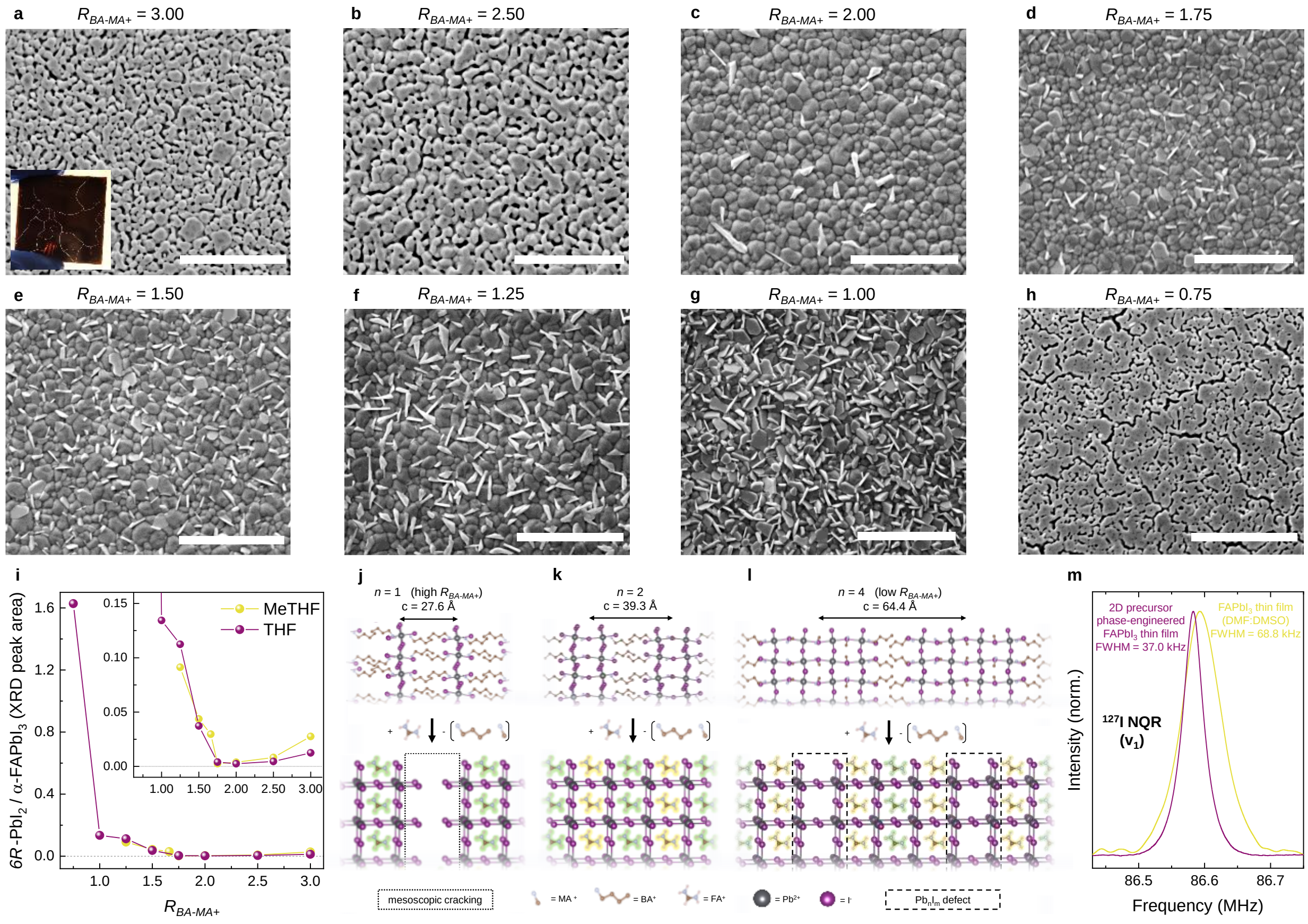

**Fig. 3 | Optimal 2D intermediate.** Scanning Electron Microscopy (SEM) images showing the microstructure of 2D-intermediate α-$FAPbI_3$ layers when a series of 2D intermediate processed from THF with varying $R_{BA\text{-}MA+}$ are converted sequentially. $R_{BA\text{-}MA+}$ = (**a**) 3.00 (inset: photograph showing macroscopic cracking of 3D perovskite layers converted from $R_{BA\text{-}MA+}$ = 3.00 intermediates due to extreme volume contraction), (**b**) 2.50, (**c**) 2.00, (**d**) 1.75, (**e**) 1.50, (**f**) 1.25, (**g**) 1.00, (**h**) 0.75. Scale bars: 5 μm. Schematic diagrams depicting the conversion of (**i**) $BA_2PbI_4$ ($n$ = 1) (**j**) $BA_2MAPb_2I_7$ ($n$ = 2), and (**k**) $BA_2MA_3Pb_4I_{13}$ ($n$ = 4) into 3D α-$FAPbI_3$. $FA^+$ highlighted in green has replaced $BA^+$ in the organic channels of the 2D intermediate, while $FA^+$ highlighted in yellow has replaced $MA^+$ in the inorganic layers. Double-headed arrows display layer thickness. (**l**), Plot showing lead iodide ($PbI_2$) content as a phase fraction in comparison to 2D-intermediate α-$FAPbI_3$ extracted from XRD diffraction patterns (**Supplementary Fig. 15**) *via* peak integration of *6R*-$PbI_2$ (001) and α-$FAPbI_3$ (100) scattering peaks. Inset shows expanded view for low-$PbI_2$ region. (**m**) $^{127}I$ nuclear quadrupole resonance (NQR) spectra of 2D-intermediate $FAPbI_3$ and $FAPbI_3$ (processed from DMF:DMSO) thin films, which have been mechanically exfoliated and powdered for measurement.

The mesocopic cracking observed in films converted from both high <*n*> and low <*n*> (low and high $R_{BA\text{-}MA+}$, respectively; **Fig. 3a-h**) is consistent with the release of local strain in the films during thermal curing. High $R_{BA\text{-}MA+}$ precursor phases possess high spacer channel density, which is expected to facilitate rapid cation exchange during conversion. However, 2D precursor films rich in the $n$ = 1 phase possess low lead halide density (density of [$PbI_6$] octahedra), and thus require significant volume contraction upon conversion to form a continuous network of 3D corner-sharing [$PbI_6$] octahedra. By this mechanism, volume contraction leads to local strain in the converted film, and thus mesoscopic cracking upon curing. We show this process schematically in **Fig. 3j**. Conversion of low $R_{BA\text{-}MA+}$ precursor phases to a 3D perovskite phase requires substantially less volume contraction. However, as shown in **Fig. 3i**, XRD analysis of converted $R_{BA\text{-}MA+}$ = 0.75 films show that X-ray scattering intensity of the *6R*-$PbI_2$ is dominant over the 3D perovskite. XRD of the same series of 2D perovskites (**Supplementary Fig. 2**) revealed that the $R_{BA\text{-}MA+}$ = 0.75 precursor phase is the only one to contain majority $n \geq 4$ RPP phases. The lead halide density of $PbI_2$ is even greater than of 3D perovskites and so conversion of a low $R_{BA\text{-}MA+}$ precursor phase to $PbI_2$ is also expected to result in substantial local strain, and thus mesoscopic cracking of the film, as observed in **Fig. 3a**.

Next, we consider the origin of $PbI_2$ in the final $FAPbI_3$ materials. For $n > 2$ precursor phases, complete extraction of $MA^+$ from, and intercalation of $FA^+$ cations into the intact haloplumbate structure requires cation migration across multiple perovskite A-sites, penetrating through unbroken layers of [$PbI_6$] octahedra – which are not dissolved by the conversion solvent, *n*-butanol – and that act as an energetic barrier to ion intercalation. Any $MA^+$ not extracted from the precursor phase during conversion is expected to volatilise during subsequent thermal curing at 180 °C, *via* mechanisms discussed in **Supplementary Note 8**. As **Fig. 3l** illustrates, such a process yields $Pb^{2+}$ and $I^-$ -rich regions in the forming 3D perovskite, and ultimately regions of crystalline $PbI_2$ upon grain ripening during thermal curing. The occurrence of such ripening is evidenced in the substantial evolution in morphology observed during high temperature curing (**Supplementary Fig. 13g-j**). This proposed mechanism is consistent with the gradual increase in $PbI_2$ content observed in converted $1.75 \geq R_{BA\text{-}MA+} \geq 0.75$ precursor phase films. To further demonstrate this effect, and the integral role spacer molecule channels play in conversion, we attempt the conversion of neat $MAPbI_3$ (representing $(BA^+)_2(MA^+)_{n-1}Pb_nI_{3n+1}$ with $n = \infty$; $R_{BA\text{-}MA+} = 0$) with the same solution conversion protocol. Absorbance spectra of this '$MAPbI_3$ precursor phase' and the material resulting from its attempted conversion show the formation of $PbI_2$ in the converted layer, and a blue-shifted absorption onset compared to α-$FAPbI_3$, confirming that complete conversion of the $MAPbI_3$ precursor phase cannot been achieved (**Supplementary Fig. 16a**). Cross-sectional SEM images of the $MAPbI_3$ precursor phase film and the converted film (**Supplementary Fig. 16b-c**) reveal the formation of significant voids at the bottom of the perovskite layer during conversion, similar to those observed in films converted from both high and low <*n*> 2D precursor phases (**Fig. 3a-h**).

In order to estimate the strain induced or released when going from the precursors phases to the 3D perovskite, we first employ thin film profilometry and simple structural considerations. Profilometer measurements indicate that the thickness of our optimum 2D precursor phase film ($R_{BA\text{-}MA+}$ = 1.5; predominantly $n$ = 2) contracts from 1,050 nm to 725 nm upon conversion to α-$FAPbI_3$; a 32% volume contraction. Calculations using the reported unit cell volume of $BA_2MAPb_2I_7$ (3,118.7 Å$^3$)[44], which contains eight $PbI_6^{4-}$ octahedra, and that of $FAPbI_3$ (256.4 Å$^3$)[45], indicate a theoretical volume contraction of approximately 34% should be observed. These data suggest that the majority of volume contraction upon conversion occurs in the vertical plane rather than laterally. This almost entirely accounts for the expected volume contraction due to the increased density of $PbI_6$ octahedra in the 3D perovskites. Thus, almost all the strain is released in this conversion process. To further confirm the absence of microstrain in our α-$FAPbI_3$ thin films, we employ $^{127}$I nuclear quadrupole resonance (NQR) spectroscopy[41,42]. As discussed in **Supplementary Note 4**, $^{127}$I NQR is highly sensitive to local symmetry distortions in the α-$FAPbI_3$ structure induced by residual strain, A-site mixing (e.g. incorporation of $MA^+$) or X-site mixing, which lead to a broadening of the NQR transition[41,42]. In comparison to α-$FAPbI_3$ thin films made by a conventional DMF:DMSO solution processing route (in the absence of any additives), 2D precursor phase-engineered α-$FAPbI_3$ thin films show a markedly narrower $^{127}$I NQR transition (**Fig. 3m**). This is compelling evidence that none of the phenomena noted above occur in our material, notably the absence of residual strain and further confirms that our α-$FAPbI_3$ is $MA^+$-free.

The mechanics of 2D precursor phase-engineered conversion and growth require significant further investigation. However, overall it is clear that, unlike intermediates featuring coordinating solvents or other lower-dimensionality materials reported to-date[36,37,46], the <*n*> > 1 RPPs reported here hold two critical advantages as sequential deposition precursor phases for 3D perovskites: an extended corner-sharing haloplumbate structure already present in the precursor phase ready to serve as a scaffold for 3D perovskite growth, as evidenced by the direct conversion of 2D perovskite into α, not δ, phase $FAPbI_3$; and built-in organic channels within the lead-halide scaffold that can facilitate rapid intercalation of a conversion solution containing A-site cations.

To investigate if our α-$FAPbI_3$ films have useful optoelectronic properties, we compare them with α-$FAPbI_3$ perovskites fabricated *via* two conventional solution processing approaches utilising DMF:DMSO solvent mixtures; with and without methylammonium chloride (MACl) additive[47]. MACl has been shown to improve perovskite material quality, stability and optoelectronic performance[47–49]. Time-resolved PL measurements reveal an order of magnitude improvement in PL lifetime for 2D precursor phase-engineered α-$FAPbI_3$ compared with MACl-free $FAPbI_3$, and a factor of three enhancement compared to MACl-additive $FAPbI_3$ (**Fig. 4a** and **Supplementary Fig. 19b**). By fitting fluence-dependence of PL decays from films of the three materials (**Supplementary Fig. 17**), we determine a reduction in the monomolecular trap-assisted recombination rate constant, implying suppression of non-radiative recombination centres. Notably, for the conventional MACl-free $FAPbI_3$, much higher excitation fluences and charge-carrier densities are required for recombination dynamics to transition from a “trap-mediated” monomolecular decay regime to intrinsic bimolecular radiative recombination[50–52]. To probe the charge carrier mobility of these materials, we perform optical-pump-terahertz-probe (OPTP) photoconductivity spectroscopy (**Supplementary Fig. 18**) measurements. We determine that the effective sum of mobilities (electron and hole) and the bimolecular recombination rate $k_2$ of charge carriers, are largely unaffected by differences in film fabrication (**Supplementary Fig. 19a**). All the $FAPbI_3$ films studied have a very high mobility of approximately 60 $cm^2\ V^{-1}\ s^{-1}$, which are some of the highest such values reported for thin film $FAPbI_3$ and attest to the generally high quality of the fabricated layers[53]. However, as a result of the highly effective suppression of trap-mediated charge-carrier recombination, for $FAPbI_3$ fabricated *via* the 2D precursor phase-engineered approach we calculate a significantly improved charge-carrier diffusion length under solar illumination conditions of 4.6 ± 0.2 μm, significantly higher than that for either $FAPbI_3$ material processed from DMF:DMSO (1.3 ± 0.1 μm and 3.0 ± 0.2 μm for neat and MACl additive-processed $FAPbI_3$) (**Supplementary Fig. 19b**).

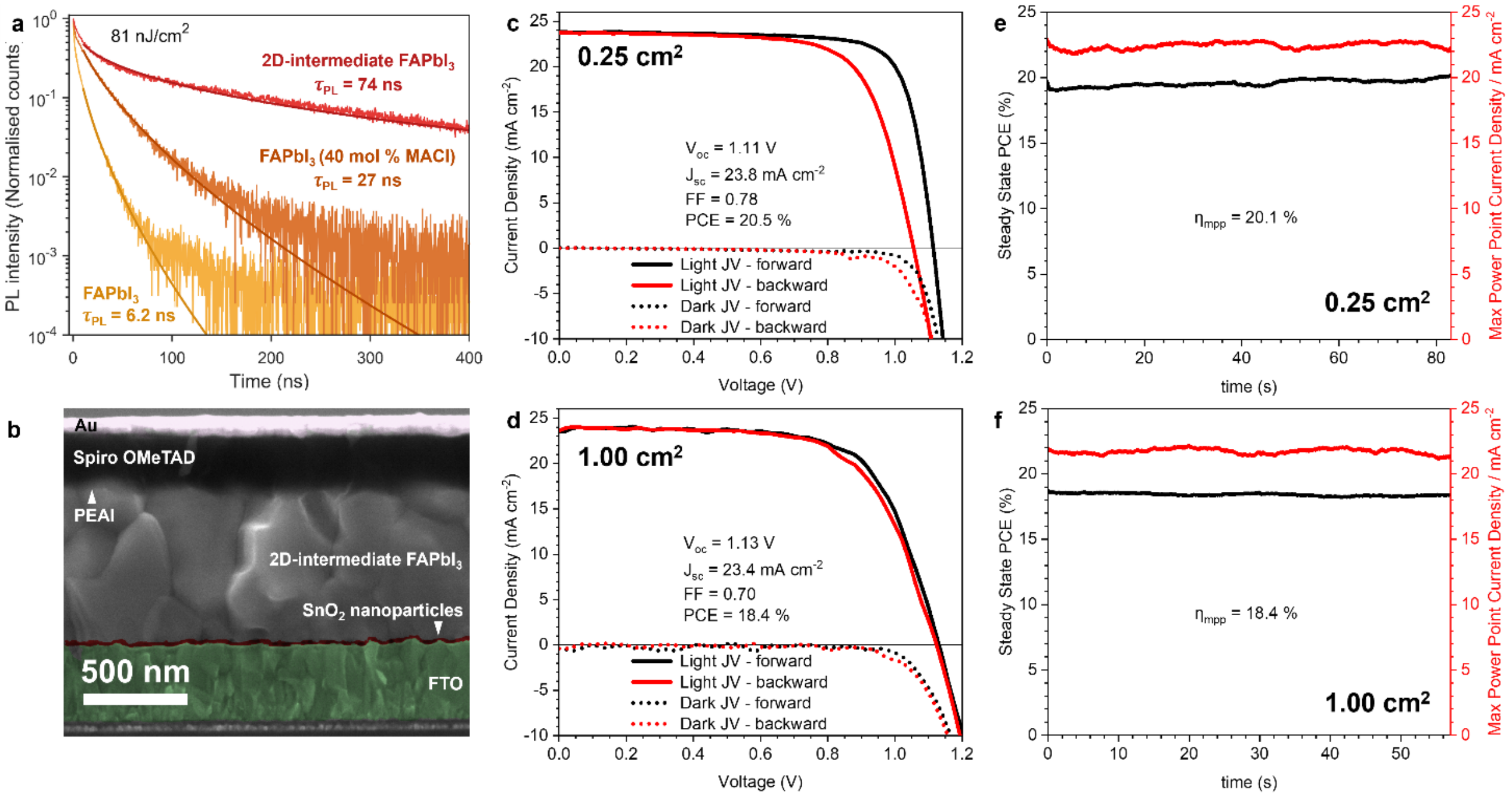

**Fig. 4 | PSC Operational Performance. a**, Time-resolved photoluminescence (PL) transients of $FAPbI_3$ materials made using different fabrication methods. PL was detected at 810 nm following pulsed-laser excitations (398 nm at 2.5 MHz repetition rate) at an excitation fluence of 81 nJ cm$^{-2}$. PL lifetimes, $\tau_{PL}$, determined by stretched exponential fits are shown on the figure labelling. Solid lines are stretched-exponential fits to the measured data. **b**, Cross-sectional scanning electron microscopy image of PSC fabricated utilising 2D precursor phase-engineered α-$FAPbI_3$. J-V characteristics and corresponding maximum power point tracked efficiency ($\eta_{mpp}$) and current density ($J_{mpp}$) measured at maximum power point for champion devices fabricated from 2D precursor phase-engineered α-$FAPbI_3$ with illuminated area 0.25 cm$^2$ (**c, e**) and 1.00 cm$^2$ (**d, f**). Light J-V and $\eta_{mpp}$ measurements were performed under simulated AM1.5 100.7 mWcm$^{-2}$ irradiance, accounting for the spectral mismatch factor. $J_{sc}$ = short-circuit current density, $V_{oc}$ = open-circuit voltage, FF = fill factor, $V_{mpp}$ = maximum power point voltage.

To determine if the improved optoelectronic properties of the 2D precursor phase-engineered α-$FAPbI_3$ films translate into improved PSCs we fabricate n-i-p structured devices composed of FTO / $SnO_2$ / α-$FAPbI_3$ / PEAI passivation / spiro-OMeTAD / Au, which we show a cross-sectional SEM image of in **Fig. 4b**. In **Fig. 4c-f** we show the current-voltage curves and maximum power point tracked efficiency ($\eta_{mpp}$) for champion devices of both 0.25 cm$^2$ and 1 cm$^2$ respectively, and determine a $\eta_{mpp}$ of 20.1 % and 18.4 %, respectively. In **Supplementary Fig. 20a** we present the distribution of optimised 0.25 cm$^2$ and 1.00 cm$^2$ PSCs. For comparison, in **Supplementary Fig. 20b-d** we present PSCs fabricated similarly but employing $FAPbI_3$ with MACl additive (40 mol%), which achieve a maximum $\eta_{mpp}$ of 18.8 %.

**Investigating α-$FAPbI_3$ stability**

Despite the favourable bandgap of α-$FAPbI_3$ for single-junction and for use as the middle junction in triple-junction PSCs, the well-known phase instability of this material is a drawback in its selection as our target composition. At room temperature the hexagonal non-perovskite *2H* (δ) phase of $FAPbI_3$ is thermodynamically preferred[54]. Although kinetic entrapment of the α-phase is possible, phase transformation is known to be accelerated by a range of external influences, including ambient humidity[4], and materials properties, such as iodide-defects[9]. In **Supplementary Fig. 21-23** we show the evolution in absorbance, microstructure (by means of SEM and visible light microscopy) and phase composition (by XRD) for the same three $FAPbI_3$ materials investigated above, over 500 hours of storage under ambient conditions (18-22 °C, 20-55% relative humidity). We find that the composition of 2D precursor phase-engineered α-$FAPbI_3$ remains almost entirely unaltered over this period, while α-$FAPbI_3$ fabricated conventionally from DMF:DMSO solvents both with and without MACl additive show substantial secondary phase formation and reduction in optical density (absorbance), particularly at the band edge, over the first 300 hours of ambient storage.

In **Supplementary Fig. 25** we show the unencapsulated "shelf stability" (stored in dry air, <10% relative humidity, at room temperature in darkness; adapted ISOS-D-1) of our n-i-p PSCs based on 2D precursor phase-engineered α-$FAPbI_3$, where we observe only minimal drop in performance over 18,000 hours (~2 years). Generally, such estimates of stability are of only limited usefulness in assessing long-term PSC stability[1]. However, in this instance, since α-$FAPbI_3$ is known to possess only kinetic metastability such sustained performance confirms the unexpectedly promising phase-stability of our material.

To investigate the stability of our α-$FAPbI_3$ material under operational conditions and explore the relationship between perovskite processing route and resultant stability we perform *in-situ* XRD tracking the accelerated thermal degradation (130 °C, ~30% RH) of 2D precursor phase-engineered α-$FAPbI_3$, (**Fig. 5a**) which we discuss in detail in **Supplementary Note 7**. We find that α-$FAPbI_3$, $MAPbI_3$ and $FA_{0.83}Cs_{0.17}Pb(I_{0.9}Br_{0.1})_3$ thin films fabricated *via* DMF:DMSO solvent systems all degrade to $PbI_2$ much more rapidly than our α-$FAPbI_3$ processed from highly volatile MeTHF, BA and BuOH solvents (**Fig. 5b-d**). Use of higher curing temperatures (180 °C) during processing reduces the rate of thermal degradation, but compromises perovskite crystallinity and phase purity (**Fig. 5e-f**). We also show that, besides 2D precursor phase-engineered α-$FAPbI_3$, remarkably the most thermally stable perovskite composition investigated is $MAPbI_3$ processed from highly volatile ACN, which is much more thermally stable than $MAPbI_3$ *via* DMF:DMSO (**Fig. 5g**). From these results, we hypothesised that poor thermal stability may be due to retention of residual processing solvents in thin films processed from low volatility DMF and DMSO. To confirm that such residual processing solvents are indeed released during thermal degradation, we employ thermal desorption-gas chromatography-mass spectrometry (TD-GCMS), whereby volatile organic compounds (VOCs) are desorbed from perovskite thin films and individually characterised. These may be VOCs retained during processing or degradation products of the perovskite material. TD-GCMS is extremely sensitive; we calculate the detection limit of DMSO retained within a 25 $cm^2$ thin film to be 0.002 wt.%[55] (**Supplementary Note 7**). We confirm that DMSO is released from α-$FAPbI_3$, $MAPbI_3$ and $FA_{0.83}Cs_{0.17}Pb(I_{0.9}Br_{0.1})_3$ films fabricated from DMF:DMSO. No solvents are released from either $MAPbI_3$ or our α-$FAPbI_3$ processed from ACN and MeTHF, respectively, confirming a direct link between reduced perovskite thermal stability and release of residual processing solvents. Equally, we detect no low volatility residual solvents in α-$FAPbI_3$ and $FA_{0.83}Cs_{0.17}Pb(I_{0.9}Br_{0.1})_3$ processed *via* DMF:DMSO and cured at 180 °C. However, we do detect release of *sym*-triazine from these materials, as is the case in the degradation of every $FA^+$-containing perovskite investigated excluding 2D precursor phase-engineered α-$FAPbI_3$ (15 different materials, **Supplementary Fig. 28**). A total of seven individual precursor phase-engineered α-$FAPbI_3$ samples were analysed, with no *sym*-triazine detected. *Sym*-triazine has been reported as a thermal degradation product of $FA^+$ [56–58]. By isolating a range of processing parameters, we identify that the 2D precursor phase-engineering crystallisation process itself is most likely responsible for the suppression of $FA^+$ volatilisation in our α-$FAPbI_3$ (see discussion in **Supplementary Note 7**). We summarise these findings in **Fig. 5h**.

To compare the stability of our α-$FAPbI_3$ to other highly stable perovskite compositions under conventional environmental stressing conditions we encapsulate perovskite films in glass-glass laminates employing an industry standard butyl-rubber edge seal, and subject the perovskite material to the conditions of two of the most demanding stability tests for perovskites; damp-heat (85 °C and 85 % relative humidity; ISOS-D-3) and light-soaking under elevated temperature (85 °C under a 1-sun equivalent light source; ISOS-L-2). Impressively, we observe very little change in the UV-vis absorption spectra for our α-$FAPbI_3$ films when aged for over 3000 hours (3 times the IEC standard) under both stressing conditions. In **Supplementary Fig. 32a-b**, we show the optical density (OD) of the films, at wavelengths between 500 to 510 nm, as a function of stressing time. This gives a relatively sensitive indication of the evolution of "pin-holes" in the perovskite films, which have previously been observed to form during degradation[3,59]. The relatively high OD of our α-$FAPbI_3$ films over the aging time, indicates that few pinholes are forming, which we confirm this with visible light microscopy, where in **Fig. 5i-l** we show images of the different films after >3,000 hours stressing.

To assess if the improved material stability translates into improved device stability, we integrate our α-$FAPbI_3$ into p-i-n structured PSCs. Although we have demonstrated >2 year "shelf stability" (**Supplementary Fig. 25**), n-i-p PSCs are known to suffer from a number of inherent instabilities under elevated moisture[60], temperature[61] and light[62], principally due to the instability of the transport materials usually employed in this architecture. However, reports of pristine $FAPbI_3$ in the more stable p-i-n configuration are unusual[63]. Furthermore, for 2D precursor phase-engineered α-$FAPbI_3$ we find that selection of the underlying hole transport material is severely limited by a combination of solvent incompatibility, high temperature processing of the perovskite and low mechanical adhesion with the as-deposited 2D precursor phase-engineered α-$FAPbI_3$ layer. After mitigating these limitations (discussed in the

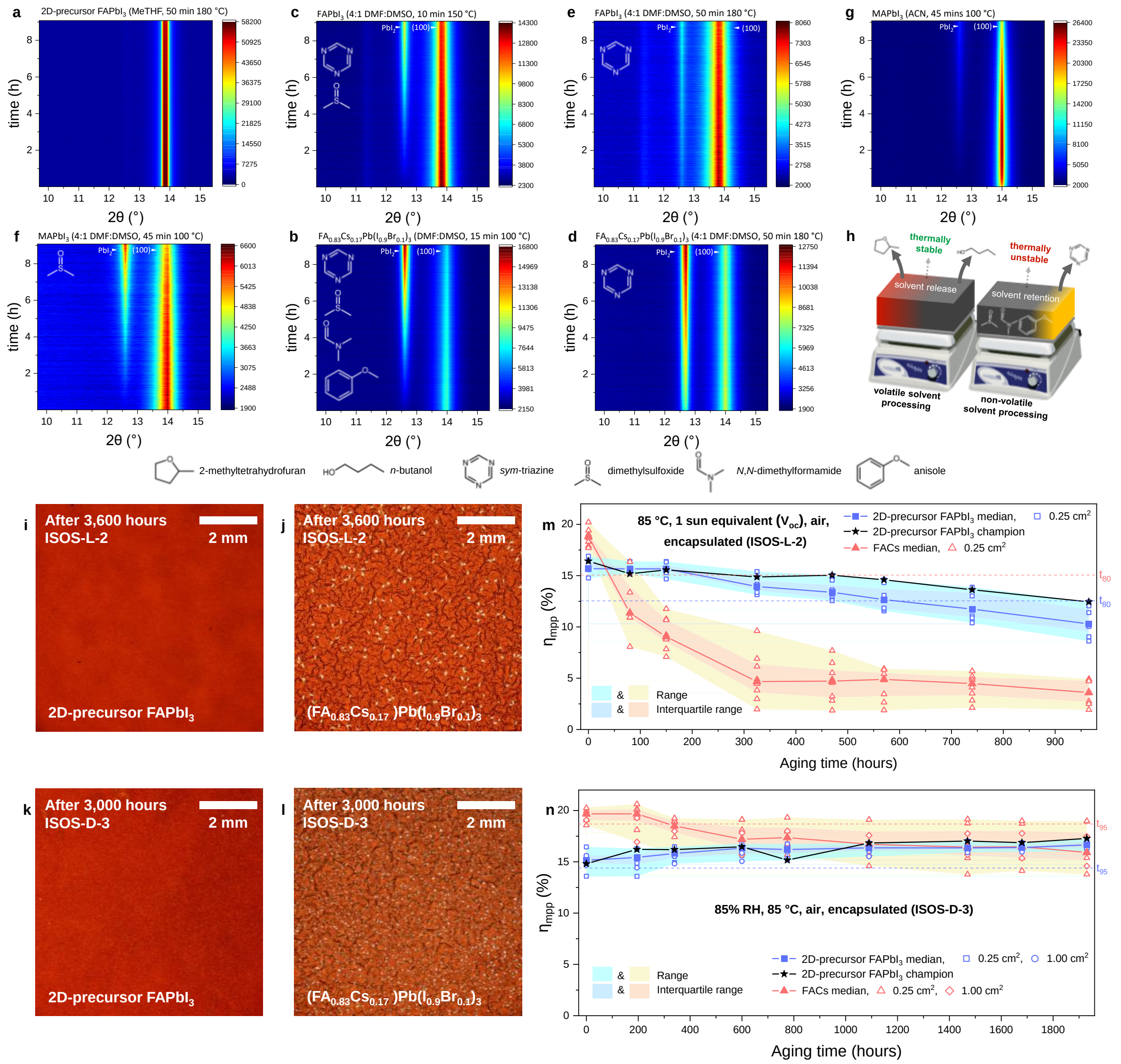

**Fig. 5 | Enhanced perovskite stability. a-g**, Contour maps depicting in-situ x-ray diffraction (XRD) analysis of perovskite materials thermally degrading whilst held at 130 °C for 9 hours. Labels indicate perovskite composition and fabrication conditions (solvent, thermal curing). Full processing conditions described in **Supplementary Note 6**. Molecular structures shown correspond to the volatile species detected during TD-GCMS analysis of each perovskite. **h**, Schematic highlighting advantages to perovskite material properties of solution processing *via* MeTHF solvent system. Visible light microscopy images of encapsulated state-of-the-art perovskite thin films after aging for 3,600 hours under 85°C, 1-sun equivalent illumination (ISOS-L-2, **i-j**) and 3,000 hours under 85 °C, 85 % relative humidity (ISOS-D-3, **k-l**) conditions. Evolution under ISOS-L-2 (**m**) and ISOS-D-3 (**n**) stressing of maximum power tracked efficiency ($\eta_{mpp}$) measured periodically on p-i-n PSCs employing 2D precursor phase-engineered $FAPbI_3$ (7 cells and 8 cells, respectively) and $FA_{0.83}Cs_{0.17}Pb(I_{0.9}Br_{0.1})_3$ (8 cells, under each condition, “FACs”) as the photoactive layer. Cells undergoing ISOS-L-2 are encapsulated with a 250 nm layer of $MoO_3$ followed by an on-cell epoxy resin and cover slip. Cells undergoing ISOS-D-3 are encapsulated in glass-glass laminates employing an industry standard butyl-rubber edge sealant (described in **Supplementary Note 10**). Dashed lines show the median $t_{80}$ $\eta_{mpp}$ values for each dataset. The median, range and interquartile range of each dataset are shown, as well as each data point within them.

Methods), we achieve substantial photovoltaic performance ($\eta_{mpp}$ = 18.8 %, **Supplementary Fig. 33**).

Exposure of these devices to the ISOS-D-2 (85 °C, dark, $N_2$) aging procedure confirms the remarkable long-term thermal stability of our α-$FAPbI_3$-based PSCs (**Fig. 5a**), retaining 95% of their initial $\eta_{mpp}$ after 1,900 hours of aging (median of 15 cells, **Supplementary Fig. 34**). Next, we isolate the effect of light-induced degradation in the absence of elevated heat (adapted ISOS-L-1). Under these conditions we again find that our α-$FAPbI_3$ PSCs are remarkably stable (projected $t_{80}$ = 3,940 hours, 21 cells, encapsulated, **Supplementary Fig. 35**). In comparison, PSCs of the same architecture employing $FA_{0.83}Cs_{0.17}Pb(I_{0.9}Br_{0.1})_3$ show reduced photostability, with a $t_{80}$ of just 850 hours (10 cells). Although our target material is α-$FAPbI_3$, we opt to compare this to an $FA_xCs_{1-x}$ perovskite since these materials have been established as having state-of-the-art stability in single-junction PSCs[2,3]. Next, we subject similar PSCs to the ISOS-L-2 test (85 °C in ambient air, full spectrum 0.76 sun illumination with cells held at open circuit) in an Atlas Suntest CPS-Plus xenon-lamp aging box, as previously for thin film materials (**Fig. 5i-j** and **Supplementary Fig. 32a-b**). We find that device degradation of PSCs based on our precursor phase-engineered α-$FAPbI_3$ achieve a promising ISOS-L-2 $t_{80}$ of 570 hours (8 cells) in comparison to $FA_{0.83}Cs_{0.17}Pb(I_{0.9}Br_{0.1})_3$-based PSCs which exhibit a $t_{80}$ of just 40 hours (7 cells) under these conditions (**Fig. 5m**). We note that this is typical for a standard $FA_xCs_{1-x}$ PSC aged under these conditions, without any stability-enhancing additives[2,3]. Our champion α-$FAPbI_3$ cell (0.25 $cm^2$) achieves a $t_{80}$ of 800 hours. Considering the finding that our α-$FAPbI_3$ thin films appear to be unchanged after 3,600 hours of aging under ISOS-L-2 (**Fig. 5i** and **Supplementary Fig. 32**), we postulated that the 570-hour $t_{80}$ measured for PSCs based on this perovskite is still limited by factors besides photoabsorber bulk stability. Careful removal of the encapsulation and n-type contact layers after 1,450 hours of ISOS-L-2 aging seems to confirm this inference. In **Supplementary Fig. 37** we present XRD patterns of the aged devices, showing minimal change in the composition of either perovskite. However, SEM images of the pristine and aged perovskite layers (**Supplementary Fig. 38**) reveal degradation-induced void formation, which is substantially more severe in the $FA_{0.83}Cs_{0.17}Pb(I_{0.9}Br_{0.1})_3$ PSCs.

Notably, these enhancements in PSC stability are achieved despite the presence of a small quantity of lead (II) iodide in the photoabsorber layer (**Fig. 3i**). *2H*-$PbI_2$ has been shown by us and others to generate defect-rich interfaces with α-$FAPbI_3$ and to undergo photodegradation to metallic $Pb^0$ and $I_2$, producing iodide vacancies. As both iodide vacancies[64] and interstitial iodide[9] defects have been implicated in accelerating the rate of α-to-δ phase transformation, the stability of our material is surprising. We note, however, that because of our high temperature processing, the $PbI_2$ phase present in our material is the *6R* polytype, which has recently been shown to form highly coherent, low-defect interfaces with α-$FAPbI_3$[65]. Moreover, as **Fig. 3a-h** clearly show, the majority of *6R*-$PbI_2$ in 2D-intermediate α-$FAPbI_3$ is in isolated grains. Thus, minimal $PbI_2$-$FAPbI_3$ interface is present in the material and any iodide vacancies generated by $I_2$ formation *in-operando* are isolated from the α-$FAPbI_3$ and so can play no part in accelerating its phase transformation.

Finally, by developing a whole-cell encapsulation strategy (**Supplementary Note 10**) p-i-n PSCs employing 2D precursor phase-engineered α-$FAPbI_3$ are able to pass the IEC61625-1 damp heat test (85 °C, 85% RH, ISOS-D-3), with >100% of initial median $\eta_{mpp}$ retained after 1,930 hours ($t_{100}$ > 1,930 hours, 8 cells, **Fig. 5n**). By contrast, comparable $FA_{0.83}Cs_{0.17}Pb(I_{0.9}Br_{0.1})_3$-based cells achieve a median $t_{95}$ of just 315 hours (8 cells). The IEC pass criterion is $t_{95}$ > 1,000 hours. Notably, the survival of the 2D precursor phase-engineered α-$FAPbI_3$ for nearly 2,000 hours under ISOS-D-3 conditions, despite the material's moisture-accelerated *2H*-phase instability, suggest that this simple three-ion perovskite may prove to be the most stable perovskite composition, matching the expectations for real world PV module deployment. To put our damp heat results into context, for copper indium gallium selenide PV modules, 3,000 hours 85 °C, 85% RH stressing has been shown to be comparable to 20 years deployment in Miami, USA[66]. We summarise all lifetimes of PSCs presented in this work in **Supplementary Table 3**.

## Conclusion

Here we have developed a volatile, low toxicity solvent system for α-$FAPbI_3$ This approach has allowed us to present an unconventional crystallisation pathway for a 3D perovskite in which the pre-existing corner-sharing $[PbI_6]^{4-}$ network present in a solid state 2D RPP perovskite is retained to direct the growth of α-$FAPbI_3$. Competitive PSC device efficiencies validate the 2D precursor phase-engineered crystallisation route, particularly considering the absence of $MA^+$ in the final 3D perovskite material and the use of a novel solvent system. We find that α-$FAPbI_3$ fabricated by our processing approach possesses substantially improved thermal and phase-stability, as compared to a broad range of state-or-the-art perovskite compositions and processing routes. We demonstrate that this improved stability is the product of several factors. Critical among these is the 2D precursor phase crystallisation route itself, which both avoids the use of low volatility solvents – which we show are often retained in perovskite thin films, leading to long-term instability – and leads to a material in which $FA^+$ degradation to *sym*-triazine under thermal stressing is supressed. Integration of our 2D precursor phase-engineered α-$FAPbI_3$ into p-i-n PSCs confirm that the enhancement in bulk perovskite stability translates to impressive device stability. In particular, the photostability under elevated temperature (ISOS-L-2) of our α-$FAPbI_3$ gives a champion $t_{80}$ lifetime of 800 hours in comparison to just 100 hours for $FA_{0.83}Cs_{0.17}Pb(I_{0.9}Br_{0.1})_3$-based PSCs. Further, we determine $t_{100}$ lifetimes of >1900 hours under damp heat stressing (ISOS-D-3). These danp heat results are particularly surprising as α-$FAPbI_3$ is considered to be metastable under ambient operational conditions. $FAPbX_3$ perovskites inherently bypass stability issues associated with A-site cation segregation and inhomogeneity[12], the possibility of creating Cs-rich impurity phases[21] and the hygroscopic nature of alkali-metal halide salts. Furthermore, by avoiding caesium, future material scarcity issues are avoided, representing a highly sustainable perovskite composition for terra watt scale PV deployment.

**Acknowledgements**

We acknowledge the contribution of Shuan Reeksting at the $MC^2$ Institute, University of Bath, for performing and advising on our TD-GCMS measurements, and Kilian Lohmann for providing the evaporated $FAPbI_3$ sample for TD-GCMS. Further, we acknowledge Diamond Light Source for time on beamline I07 under proposal SI33462-1 for the GIWAXS measurements and thank Daniel Toolan, Rachel Kilbride and Jonathan Rawle for assistance with these measurements. This project was supported by the Plastic Electronics Centre for Doctoral Training and has received funding from the European Union's Horizon 2020 research and innovation programme under the Marie Skłodowska-Curie grant agreement No 764787 and the EPSRC under grant number EP/V027131/1. I.L. thanks the AiF project (ZIM-KK5085302DF0) for financial support.

**Author Contributions**

B.M.G. conceived the experiments with support from P.H. and J.A.S. B.M.G. developed the ether-amine solvent systems, carried out characterisation of the 2D perovskites, investigation of the mechanism of 2D-intermediate conversion, fabrication of the perovskite solar cells, TD-GCMS analysis and all materials and device stability measurements besides those involving in-situ XRD measurements. S.C. assisted with fabrication of p-i-n configuration solar cells. B.M.G., P.H. and J.A.S. carried out XRD analysis. K.A.E. carried out all OPTP and TCSPC measurements and associated calculations, supervised by L.M.H. J.A.S., B.M.G. and F.Y. conducted the GIWAXS characterisation and analysis. A.S. analysed the SED data, supervised by N.K.N. J.M.B. constructed the setup used for EQE measurements and developed the software used to measure and analyse all PSC device performance. M.G.C. designed and implemented the solar simulator used for all device performance measurements. I.L. carried out and analysed the modulated SPV measurements. H.J.S supervised the project, developed the ideas and concepts with B.M.G, and raised the funds for the work. B.M.G. wrote the manuscript and P.H., J.A.S. and H.J.S. edited it. All authors contributed to proofing of the manuscript and writing/editing sections related to their contributions.

**Data Availability**

All relevant data are provided in the figures, tables and Supplementary Information. The raw data acquired from all characterization methods are available on Oxford University Research Archive.

## Competing Interests

H.J.S. is a co-founder and Chief Scientific Officer of Oxford PV Ltd., a company industrialising perovskite PV. The other authors declare no competing interests.